\documentstyle[12pt,epsfig,color]{article}
\begin{document}

\begin{center}
{\Large\bf Spherical collapse in reconstructed dark energy models}
\\[15mm]
Ankan Mukherjee\footnote{Email: ankan.ju@gmail.com}
\vskip 0.5 cm
{\em Department of Physics, Bangabasi College, \\Kolkata - 700009, India.\\
 Centre for Theoretical Physics, Jamia Millia Islamia,\\ New Delhi - 110025, India.}\\[15mm]
\end{center}

\pagestyle{myheadings}
\newcommand{\be}{\begin{equation}}
\newcommand{\ee}{\end{equation}}
\newcommand{\bea}{\begin{eqnarray}}
\newcommand{\eea}{\end{eqnarray}}

\begin{abstract}
The present work deals with the spherical collapse of matter overdensity for two reconstructed dark energy models. One of the models is reconstructed from parametrization of effective or total equation of state of energy components in the universe ($w_{eff}$ model) and the other model is the constant dark energy equation of state model, namely the $w$CDM model. The linear and nonlinear evolution of matter density contrast are studied for the present models. It is observed that the linear and even the nonlinear evolutions of density contract are almost indistinguishable in these two models.   
The critical density contrast at collapse as a function of redshift is also studied. The nature of critical density contrast is also found to be degenerate in the present models.  Further the number count of collapsed objects or dark matter halos along redshift are  also studied. Two different halo mass functions, namely the Press-Schechter mass function and the Sheth-Tormen mass function, are adopted in this context. It is observed that for both the mass functions the $w_{eff}$ model has slightly higher number of halos at very low redshift ($z<0.5$) and at higher redshift the $w$CDM model has higher number of dark matter halos. On the other hand, it is observed that the Press-Schechter mass function produces slightly higher number of dark matter halos at low redshift compared to Sheth-Torman mass function and the number of halos is higher for Sheth-Torman mass function at redshift $z>0.9$ for both the dark energy models. The results clearly show that these two highly degenerate dark energy models are distinguishable in the study of spherical collapse and galaxy cluster number counts.  
\end{abstract}


\vskip 1.0cm

\section{Introduction}

The discovery of cosmic acceleration \cite{Riess:1998cb,Perlmutter:1998np,Schmidt:1998ys} in late nineties has changed our understanding about the dynamics of the universe. The phenomenon is further confirmed by the latest cosmological observations \cite{Suzuki:2012,Crocce:2015xpb,Agathe:2019vsu,Planck:2018vyg}. The remarkable discovery has opened up a new arena of research in the field of cosmology.
The observed accelerated expansion is very astonishing as our prior knowledge suggests that the expansion should decelerate due to the attractive nature of gravitational force. Different theoretical prescriptions are proposed to explain the phenomenon of cosmic acceleration. Within the regime of General Relativity (GR), it could be explained by introducing an exotic component in the energy budget of the universe. This exotic component is dubbed as {\it dark energy}. The alleged acceleration is generated due to the effective negative pressure of dark energy. The other way to look for a possible solution of the puzzle is to modify the theory of gravity.
With the unprecedented technological advancement in observational cosmology, different cosmological parameters are constrained to a very high level of precision \cite{Planck:2018vyg,DES:2021bwg,DiValentino:2019qzk}. But we are still far away from the knowledge about actual physical entity of dark energy. Various theoretical prescriptions are there in literature regarding the possible candidates dark energy. Comprehensive reviews in this direction are there in \cite{Copeland:2006wr,Sahni:2006pa}.

All the viable dark energy models produce very similar background cosmological evolution. The desired evolution can be achieved from the reconstruction of the dark energy equation of state  ($w_{DE}$) or other dark energy parameters. Most of the models generate cosmic evolution identical to the {\it cosmological constant} ($\Lambda$CDM) model for which $w_{DE}=-1$.  To distinguish the models, it is necessary to study the evolution of perturbation, specially at non-linear level. Non-linear evolution of matter perturbation is the fundamental process behind the formation of dark matter halos and large scale structures in the universe. The abundance of dark matter halos is directly related to the underlying cosmological model \cite{Sunyaev:1980,Majumder:2004cs,Diego:2004uw,Fang:2006dt}. In particular, the formation  of large scale structure in the universe is sensitive to the normalized matter power spectrum ($\sigma_8$), matter density parameter ($\Omega_{m}$) and the dark energy equation of state ($w_{DE}$) \cite{Holder:2001db}.

A simple and highly useful approach to study the evolution of matter perturbation is the spherical collapse model \cite{Gunn:1972sv,Liddle:1993fq,Padmanabhan:1999}. This semi-analytic approach is adopted in  the present work to study the non-linear evolution of matter overdensity and formation of large scale structure. The basic idea is to study the evolution of spherical overdensity using the fully nonlinear equation derived from Newtonian hydrodynamics. The overdence region is assumed to be spherically symmetric and to have a uniform density which is higher than the background density of dark matter. It is considered as a closed sub-universe expanding with Hubble flow. But the expansion slows down and after reaching a maximum radius, it stars compression and eventually collapses due to gravitational attraction. A virilization of gravitational potential and thermal energy due to random motion of matter particles is introduced in this context to explain the finite size of the collapsed objects. 
The effect of dark energy on the clustering of dark matter can be probed in the spherical collapse model. Several studies in this direction are there in literature \cite{Mota:2004pa,Nunes:2005fn,Mota:2008ne,He:2010ta,Pace:2010,Pace:2013pea,Basse:2010,Wint:2010,Devi:2010qp,Delliou:2012ik,Nazari-Pooya:2016bra,Setare:2017,Sapa:2018jja,Rajvanshi:2018xhf,Rajvanshi:2020das,Barros:2019hsk,Pace:2019vrs,Ruchika:2020avj}. Another numerically sophisticated  approach to study the nonlinear evolution of cosmological perturbations and formation of large scale structures in the universe is the N-body simulation \cite{Jennings:2012,Maccio:2003yk,Baldi:2010vv,Boni:2011}.

As already mentioned, the formation of large scale structure in the universe is effected by underlying dark energy model through the non-linear evolution of matter density and the associated power spectrum. Another quantity of interest in the study of  spherical collapse model is the critical density contrast ($\delta_c(z_c)$). It is defined as the value of linear density contrast at a redshift ($z_c$) where the nonlinear density contrast diverges. Critical density contrast is an important quantity in the expression of mass function which is the most important tool in the study of cosmic large scale structure. The objects, formed due to the collapse of dark matter overdensity, are called the dark matter halos. The galaxy clusters are embedded in the dark matter halos due to gravitational attraction. Thus the  distribution of galaxy clusters follows the same distribution as the dark matter halos. Thus the observed distribution of galaxy cluster is the probe of dark matter distribution in the universe.

In the present work, the spherical collapse of matter overdensity is studied for two different reconstructed dynamical dark energy models. One is the wCDM model {\it (constant dark energy equation of state with cold dark matter)} and the other one is reconstructed from parametrization of total or effective equation of state of the energy components in the universe \cite{Mukherjee:2016eqj}.

The paper is presented as the following. In section \ref{model}, the reconstruction of the dark energy models and the observational constraints on the equation of states are discussed. Linear and nonlinear evolution of matter overdensity under spherical collapse scenario and the nature of  critical density at collapse are studied in section \ref{sphericalcollapse}. In section \ref{clustercount} the number count of collapsed objects or dark matter halos for the dark energy models are studied using two different halo mass function formulas. Finally, in section \ref{conclu}, it is concluded with overall discussion about the results.

\section{Reconstructed dark energy models}
\label{model}

The present trend of studying properties of dark energy is reconstruction which is a reverse engineering of the model. Two types of approaches are there in literature for reconstruction of dark energy model. One is parametric approach, where a parametric form of any cosmological quantity, like the dark energy equation of state, are assumed and the model parameters are constrained from observational data using statistical techniques. The other approach is nonparametric, where the evolution of any cosmological quantity is constructed from the observational  data using statistical techniques without assuming any parametric form.   

The dark energy models, adopted in the present study, are obtained from parametric reconstruction. One is based on the parametrization of total or effect equation of state discussed in \cite{Mukherjee:2016eqj}. The other one is the wCDM dark energy model, where the dark energy equation of state is assumed to be a constant. It is one of the extensively studied phenomenological model in the context of late-time dynamics of the universe. The basic motivation of wCDM model is to study the dynamics when the equation of state is allowed to have value different from $-1$ which is the value of dark energy equation of state parameter for {\it cosmological constant} model of dark energy. 

In the present study, these two models are adopted as both these models are based on simple parameterizations and produce identical cosmological dynamics \cite{Mukherjee:2016eqj}. They are indistinguishable even at kinematical statefinder parameter space \cite{Mukherjee:2016eqj}. Besides, these two models are purely phenomenological. There is no prior assumptions about any particular type of dark energy or about the physical entity of dark energy. Though there are some other phenomenological parameterizations of dark energy models like the CPL ( Chevallier-Polarski-Linder) \cite{Chevallier:2000qy,Linder:2002et}, GCG (generalized Chapligin gas) parametrization \cite{Sen:2005sk}. JBL (Jassal-Bagla-Padmanabhan) parametriztion \cite{Jassal:2004ej}, PEDE (Phenomenological Emergent Dark Energy) \cite{Li:2019yem} etc. which are normally proposed based on specific choices of dark energy candidates. The equation of state in those models differ significantly from each other and the values do not remain constant. On the other hand, it is shown in \cite{Mukherjee:2016eqj} that the equation of state parameter and the associated uncertainty for wCDM and $w_{eff}$ model behave in very similar fashion.  Now if the present study of nonlinear density perturbation and cluster number count can successfully break the degeneracy of these two models, it would be identified as a powerful method to distinguish  dark energy models.

In a spatially flat FLRW (Friedmann-Lemaitre-Robertson-Walker) space-time, the Friedmann equations, yield from the Einstein's field equation, are
\be
3H^2=8\pi G\rho_{tot},
\ee
\be
2\dot{H}+3H^2=-8\pi G p_{tot},
\ee
where $\rho_{tot}$ and $p_{tot}$ are the total energy density and the total pressure-like contribution of all the components in the universe.  The total or effective equation of state is defined as,
\be
w_{eff}=\frac{p_{tot}}{\rho_{tot}}.
\ee
The components in the energy budget of the universe are matter, dark energy and radiation. The total energy density can be written in terms of the energy densities of different components as,
\be
\rho_{tot}= \rho_{m} + \rho_{DE} + \rho_{r},
\ee
where $\rho_{m}$ is the matter energy density, $\rho_{DE}$ is the dark energy density and $\rho_{r}$ is the radiation energy density. Similarly $p_{tot}$ can be written as the summation of the pressurelike contributions from all the components as,
\be
p_{tot}= p_{m} + p_{DE} + p_{r}.
\ee
The dark energy equation of state parameter is defined as, $w_{\footnotesize DE}=\frac{p_{\footnotesize DE}}{\rho_{\footnotesize DE}}$. The matter component is considered to be the dust matter for which $p_m=0$

As already mentioned that one of the models, adopted in the present context, is reconstructed from a parametric form of the $w_{eff}$ \cite{Mukherjee:2016eqj}. The phenomenological expression of $w_{eff}$ is introduced in \cite{Mukherjee:2016eqj} as,
\be
w_{eff}(z)=-\frac{1}{1+\alpha(1+z)^n},
\ee
where $z$ is the redshift, and ($\alpha,n$) are the model parameters. The expression of Hubble parameter is obtained for this model as \cite{Mukherjee:2016eqj},
\be
H(z)=H_0\Bigg(\frac{1+\alpha(1+z)^n}{1+\alpha}\Bigg)^{\frac{3}{2n}},
\label{hubbleparam1}
\ee
where $H_0$ is the present value of the Hubble parameter. The dark energy equation of state for the model is expressed as \cite{Mukherjee:2016eqj},
\be
w_{DE}(z)= -\frac{\Bigg(\frac{1+\alpha(1+z)^n}{1+\alpha}\Bigg)^{\frac{3}{n}}-\Big(\frac{\alpha}{1+\alpha}\Big)(1+z)^n\Bigg(\frac{1+\alpha(1+z)^n}{1+\alpha}\Bigg)^{\frac{3}{n}-1}}{\Bigg(\frac{1+\alpha(1+z)^n}{1+\alpha}\Bigg)^{\frac{3}{n}}-\Big(\frac{\alpha}{1+\alpha}\Big)^{\frac{3}{n}}(1+z)^3}.
\ee

On the other hand, the expression of Hubble parameter $H(z)$ for wCDM model is given as,
\be
H(z)=H_0\Bigg( \Omega_{m0}(1+z)^3 + (1-\Omega_{m0})(1+z)^{3(1+w)}  \Bigg)^{\frac{1}{2}},
\label{hubbleparam2}
\ee
where $\Omega_{m0}$ is the present matter density parameter and $w=w_{DE}$, a constant in this model.

\begin{figure}[tb]
\begin{center}
\includegraphics[angle=0, width=0.41\textwidth]{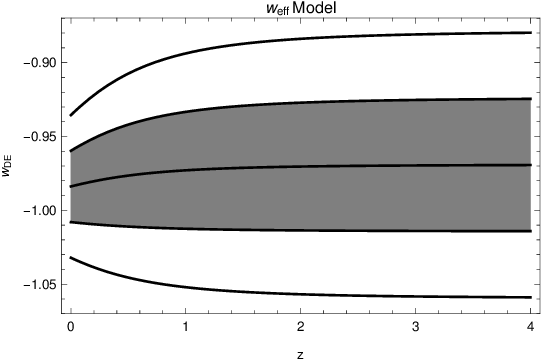}
\includegraphics[angle=0, width=0.41\textwidth]{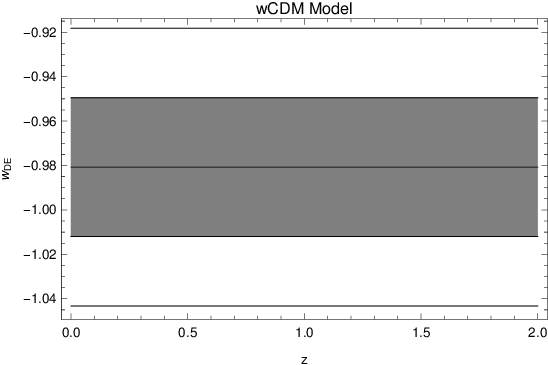}
\includegraphics[angle=0, width=0.4\textwidth]{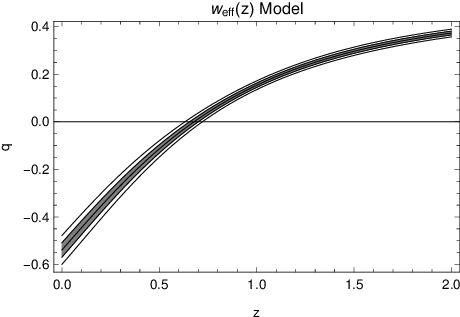}
\includegraphics[angle=0, width=0.4\textwidth]{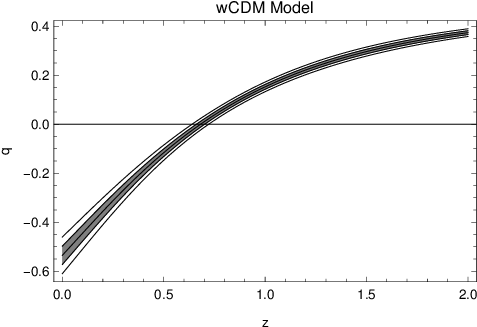}
\end{center}
\caption{{\small Plots of dark energy equation of state parameter ($w_{DE}$) (upper panels) and the deceleration parameter (lower panels) as functions of redshift $z$ for $w_{eff}$ model and $w$CDM model. The 1$\sigma$ and  2$\sigma$ confidence regions along with the best fit curves are shown \cite{Mukherjee:2016eqj}.}}
\label{wDEweffplot}
\end{figure}

In \cite{Mukherjee:2016eqj}, the constraints on the model parameters are obtained by maximum likelihood  analysis using different observational data sets, namely the observational Hubble parameter data, baryon acoustic oscillation data, distance modulus data of type-Ia supernovae and cosmic microwave background (CMB) distance prior measurement. The values of the parameters  for $w_{eff}$ model are obtained as $\alpha=0.444\pm 0.042$, $n=2.907\pm 0.136$ at 1$\sigma$ and  similarly the values of the parameters for wCDM model are obtained as $w=-0.981\pm 0.031$, $\Omega_{m0}=0.296\pm 0.011$ at 1$\sigma$ \cite{Mukherjee:2016eqj}. The evolution of dark energy equation of state parameter ($w_{DE}$) and the deceleration parameter $(q)$ for these two reconstructed models are shown in figure \ref{wDEweffplot}. In the $w_{eff}$ model the matter density parameter $\Omega_{m0}$ can be expressed in terms of the model parameter $\alpha$ and $n$. From the expression of Hubble parameter for $w_{eff}$ model (equation \ref{hubbleparam1})) it is clear that $H^2(z)/H^2_0$ can be expressed as a power series. In the series expansion of the right hand side, there is a term with $(1+z)^3$ which is actually the matter density term. The coefficient of the $(1+z)^3$ term is identified as the present matter density parameter ($\Omega_{m0}$). The coefficient of $(1+z)^3$ term  in the series expansion is $\left(\frac{\alpha}{1+\alpha}\right)^{\frac{3}{n}}$.  So the matter density parameter for the reconstructed $w_{eff}$ model is expressed as $\Omega_{m0}=\left(\frac{\alpha}{1+\alpha}\right)^{\frac{3}{n}}$ and thus the value is obtained as  $\Omega_{m0}=0.296\pm 0.011$ at 1$\sigma$ \cite{Mukherjee:2016eqj} which is exactly same to value obtained in the analysis for $w$CDM. The $w$CDM is equivalent to the cosmological constant model ($\Lambda$CDM) for $w=-1$. On the other hand the $w_{eff}$ model mimics the $\Lambda$CDM for parameter value $n=3$. Both the models are found to be consistent with the $\Lambda$CDM at 1$\sigma$ confidence level. As already mentioned, these two reconstructed dark energy models show highly degenerate evolution of cosmological background \cite{Mukherjee:2016eqj}. In the present work, the linear and non-linear evolutions of matter perturbation and formation of dark matter halos are studied for these two models. The parameter values, obtained in  \cite{Mukherjee:2016eqj}, are utilized in the present work.

\section{Matter density perturbation and spherical collapse}
\label{sphericalcollapse}

\begin{figure}[htb]
\begin{center}
\includegraphics[angle=0, width=0.284\textwidth]{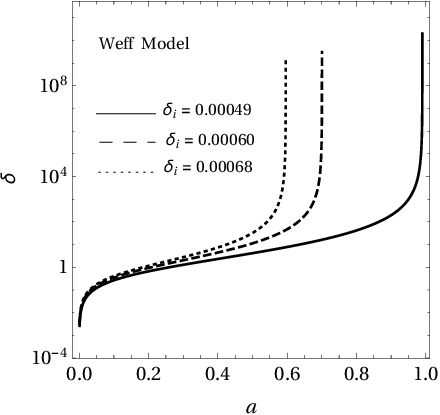}
\includegraphics[angle=0, width=0.284\textwidth]{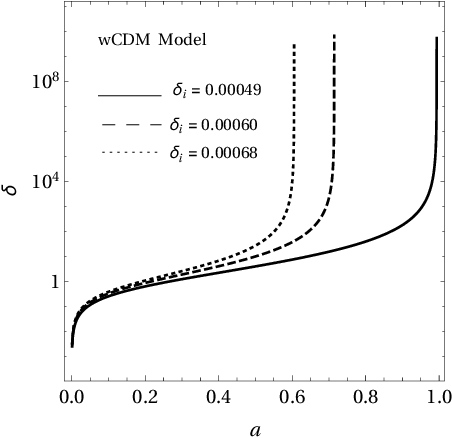}\\
\includegraphics[angle=0, width=0.284\textwidth]{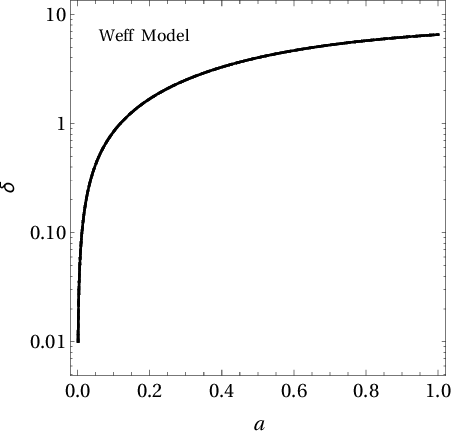}
\includegraphics[angle=0, width=0.284\textwidth]{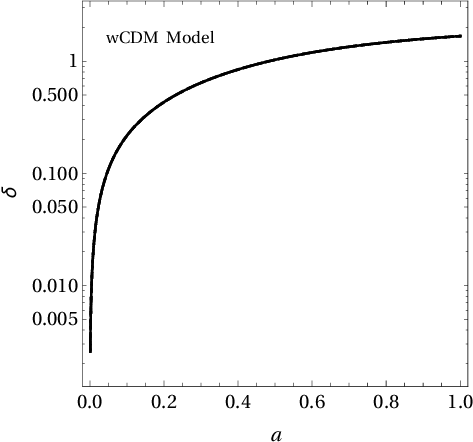}
\end{center}
\caption{{\footnotesize Evolutions of matter density contrast $\delta(a)$ for the reconstructed $w_{eff}$ model (left panels) and $w$CDM model (right panels). The upper panels show the nonlinear evolution of $\delta$ and lower panels show the linear evolution of $\delta$.
The plots are obtained for three different initial values of $\delta$ (mentioned in the figure), fixed at $a_i=0.00001$ and $\delta'(a_i)=0.001$. The values of the parameters are fixed at $\alpha=0.444$ and $n=2.907$ for $w_{eff}$ model and $w=−0.981$ and $\Omega_{m0}=0.296$ for $w$CDM. The curves of linear $\delta$ are overlapping for these three initial conditions (lower panels). }}
\label{fig:dela_plots}
\end{figure}

\begin{figure}[h]
\begin{center}
\includegraphics[angle=0, width=0.35\textwidth]{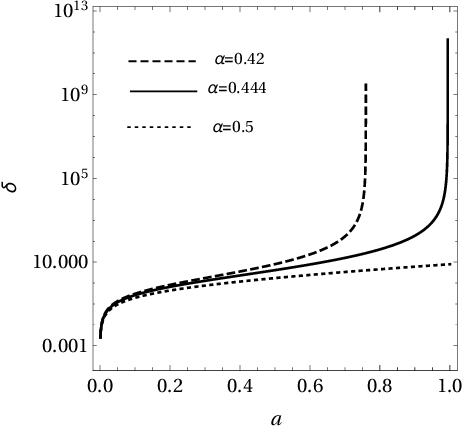}
\includegraphics[angle=0, width=0.35\textwidth]{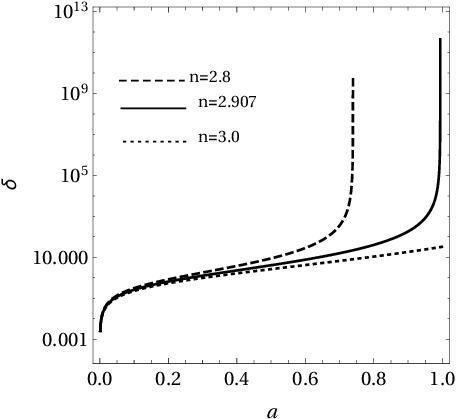}
\end{center}
\caption{{\small Nonlinear evolution of $\delta$ for the reconstructed $w_{eff}$ model for different values of the model parameters. The left panel shows  the nonlinear evolution of $\delta$ for different values of parameter $\alpha$ (fixing the value of at the best fit $n=2.907$) and the right panel shows the  nonlinear evolution of $\delta$ for different values of the parameter $n$ (fixing the value of at the best fit $\alpha=0.444$). The initial conditions to solve the equation (\ref{del_nonlin_a}) are fixed at scale factor $a_i=0.00001$ as $\delta_i=0.00049$ and $\delta'_i=0.001$. }}
\label{fig:NL_dela_weff}
\end{figure}
\begin{figure}[h]
\begin{center}
\includegraphics[angle=0, width=0.35\textwidth]{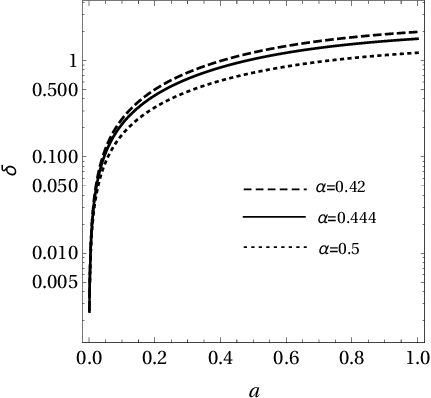}
\includegraphics[angle=0, width=0.35\textwidth]{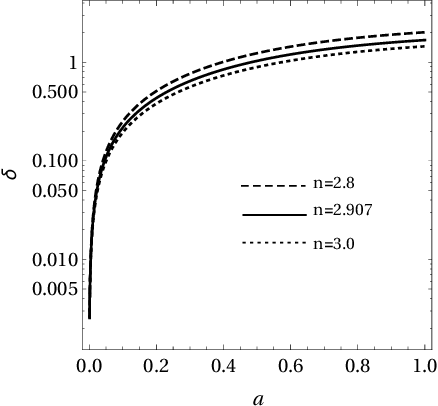}
\end{center}
\caption{{\small Linear evolution of $\delta$ for the reconstructed $w_{eff}$ model for different values of the model parameters. The left panel shows  the linear evolution of $\delta$ for different values of parameter $\alpha$ (fixing the value of at the best fit $n=2.907$) and the right panel shows the  linear evolution of $\delta$ for different values of the parameter $n$ (fixing the value of at the best fit $\alpha=0.444$). The initial conditions to solve the equation (\ref{del_lin_a}) are fixed at scale factor $a_i=0.00001$ as $\delta_i=0.00049$ and $\delta'_i=0.001$.  }}
\label{fig:L_dela_weff}
\end{figure}

\begin{figure}[h]
\begin{center}
\includegraphics[angle=0, width=0.35\textwidth]{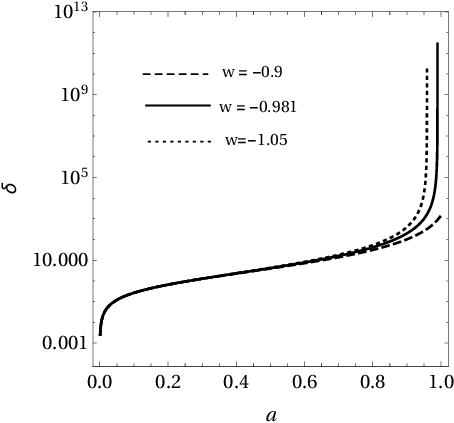}
\includegraphics[angle=0, width=0.35\textwidth]{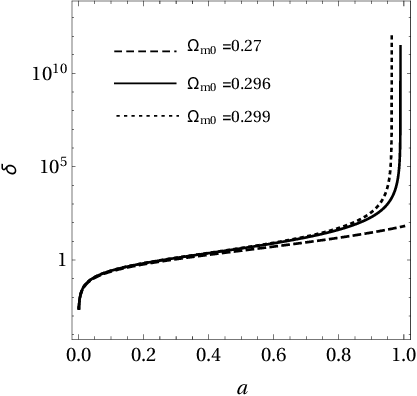}
\end{center}
\caption{{\small Nonlinear evolution of $\delta$ for the reconstructed $w$CDM model for different values of the model parameters. The left panel shows  the nonlinear evolution of $\delta$ for different values of parameter $w$ (fixing the value of $\Omega_{m0}=0.296$, bets fit) and the right panel shows the  nonlinear evolution of $\delta$ for different values of the parameter $\Omega_{m0}$ (fixing the value of $w$ at the best fit $w=-0.981$) . The initial conditions to solve the equation (\ref{del_nonlin_a}) are fixed at scale factor $a_i=0.00001$ as $\delta_i=0.00049$ and $\delta'_i=0.001$.}}
\label{fig:NL_dela_wCDM}
\end{figure}

The evolution of background cosmology is studied assuming the homogeneity and isotropy of the universe. But to understand the formation of large scale structure in the universe, it is essential to study the evolution of perturbation in the matter component. It is convenient to represent the matter perturbation in terms of density contrast, defined as  $\delta=\Delta\rho_m/\rho_m$ where $\Delta\rho_m$ is the deviation from homogeneous matter density $\rho_m$. The size of the overdense region increases due to Hubble expansion. Besides, it gathers mass from the surrounding due to gravitational attraction. The increasing gravitational force acts opposite the to the Hubble expansion and at certain time the expansion stops. Then the overdense region starts to collapse due to extremely strong gravitational force. It is called the gravitational collapse which is the fundamental process behind the formation of large scale structure in the universe. We need to study the non-linear evolution of matter perturbation to understand the formation of large scale structure. Spherical collapse \cite{Gunn:1972sv,Liddle:1993fq,Padmanabhan:1999} is the simplest model to study the gravitational collapse and structure formation in the universe. It is a semi-analytic approach. In this model the overdense region is assumed to be spherically symmetric. The matter density contrast evolves with time according to the non-linear differential equation given as \cite{Pace:2010},
  
\be
\ddot{\delta}+2H\dot{\delta}-4\pi G\rho_m\delta(1+\delta)-\frac{4}{3}\frac{\dot{\delta}^2}{1+\delta}=0.
\label{del_nonlin}
\ee
Equation (\ref{del_nonlin}) can be linearized by neglecting the higher order terms of $\delta$. The linear differential equation (\ref{del_nonlin}) is given as,
\be
\ddot{\delta}+2H\dot{\delta}-4\pi G\rho_m\delta=0.
\label{del_lin}
\ee
Using scale factor $a$ as the argument of differentiation in equation (\ref{del_nonlin}) yield,  
\be
\delta''+\left(\frac{h'}{h}+\frac{3}{a}\right)\delta'-\frac{3\Omega_{m0}}{2a^5h^2}\delta(1+\delta)-\frac{4}{3}\frac{\delta'^2}{(1+\delta)}=0.
\label{del_nonlin_a}
\ee
Similarly the linear equation (eq. \ref{del_lin}) with scale factor $a$ as the argument of differentiation  is  given as, 
\be
\delta''+\left(\frac{h'}{h}+\frac{3}{a}\right)\delta'-\frac{3\Omega_{m0}}{2a^5h^2}\delta=0.
\label{del_lin_a}
\ee
Here the prime denotes the differentiation with respect to $a$ and the $h$ is the scaled Hubble parameter $h(a)=H(a)/H_0$.
Equation (\ref{del_nonlin_a}) and (\ref{del_lin_a}) are studied numerically for the reconstructed $w_{eff}$ model and $w$CDM model. It is assumed in the present analysis that the dark energy is homogeneously distributed without any clustering. In the present study, similar methodology as in \cite{Ruchika:2020avj} is adopted.


The initial conditions for the numerical solutions of equation (\ref{del_nonlin_a}) and (\ref{del_lin_a}) are fixed at $a_i=0.00001$. The radiation energy density term is incorporated in the expression of the Hubble parameter in the numerical analysis as the contribution of radiation was also significant where the initial condition is fixed.   Figure \ref{fig:dela_plots} shows the linear and nonlinear evolutions of $\delta(a)$ for these two models. In figure \ref{fig:dela_plots} the plots are obtained for three different initial values of $\delta$ keeping the initial value of $\delta'$ same. Figure \ref{fig:dela_plots} shows that the nonlinear density contrast diverges very close to scale factor $a=1$ for the boundary value $\delta(a_i)=0.00049$, $\delta'(a_i)=0.001$ and the collapse occurs early when the initial value of density contrast is increased. It is interesting to note that the linear evolution of $\delta$ is identical for these three initial conditions and hence the linear curves are overlapping (lower panels figure \ref{fig:dela_plots}). 
 Both the models show very similar evolution of linear and nonlinear matter density contrast. The effects of the model parameter values on the linear and nonlinear evolution of $\delta$ are also investigated in the present context. In figure \ref{fig:NL_dela_weff} the nonlinear evolution of $\delta$ is investigated for $w_{eff}$ model with  different values of the model parameters $\alpha$ and $n$. The initial conditions to solve the nonlinear evolution equation (eq.(\ref{del_nonlin_a})) are fixed at $a=0.00001$ and the initial values are $\delta_i=0.00049$ and $\delta'_i=0.001$. As this initial values produce the nonlinear collapse very close to scale factor $a=1$ for the best fit values of the parameters, it would be easier to study the sensitive dependence of nonlinear collapse on the model parameter values. It is observed from the plots that the lower value of parameter $\alpha$ allows the $\delta$ to collapse earlier and with the increase of value of $\alpha$ the collapse gets delayed. Even we can see that for $\alpha=0.5$, the collapse does not occur within the present time. The value of parameter $n$ is fixed at the best fit value obtained in the analysis. The right panel of the figure shows the nonlinear-$\delta$ curves for different values of the parameter $n$, keeping the value of $\alpha$ fixed at the best fit. The curves show that the collapse happens earlier in case of lower value of $n$ and the occurrence of the collapse get delayed with the increase in the value of $n$. The plots clearly reveal the sensitive dependence of nonlinear evolution of $\delta$ on the values of the model parameters. Figure \ref{fig:L_dela_weff} shows the linear evolution of $\delta$ for the $w_{eff}$ model for different values of the parameters. The initial condition to solve the equation (\ref{del_lin_a}) are fixed at $a=0.00001$ and the initial values are $\delta_i=0.00049$ and $\delta'_i=0.001$, the same initial conditions fixed to solve the nonlinea equation also. The left panel of figure\ref{fig:L_dela_weff} shows the linear-$\delta$ curves for different values of $\alpha$ and the right panel shows the same for different values of parameter $n$. It is clear from the plots in figure \ref{fig:L_dela_weff} that the linear evolution of $\delta$ is less effected by the variation of the model parameter values.
The effect of variation of parameter values on the nonlinear-$\delta$ evolution is also studied for the $w$CDM model. In figure \ref{fig:NL_dela_wCDM} the nonlinear-$\delta$ curves are shown for $w$CDM model. The initial conditions are same as it was in case of $w_{eff}$ model. The left panel shows plots for different values of the parameter $w$ and right panel shows the nonlinear-$\delta$ curves for different values of the parameter $\Omega_{m0}$. The plots on the right panel shows that collapse occurs earlier in case of higher value of the matter density parameter $\Omega_{m0}$. On the other hand, the left panel reveals that the collapse occurs early in case of lower value of dark energy equation of state $w$. The linear evolution of $\delta$ for $w$CDM model again shows no significant variation with the change in parameter values, though the plots are not shown here. 

\begin{figure}[tb]
\begin{center}
\includegraphics[angle=0, width=0.36\textwidth]{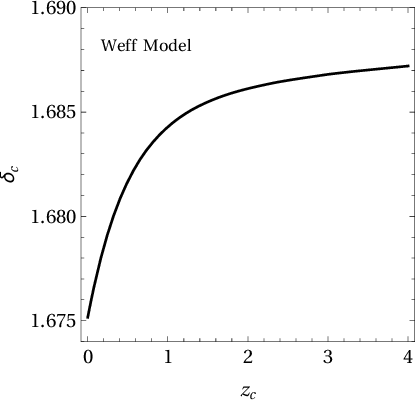}
\includegraphics[angle=0, width=0.36\textwidth]{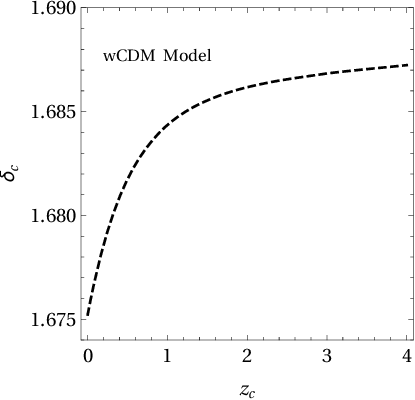}
\end{center}
\caption{{\small Critical density at collapse ($\delta_c$) as a function of redshift. The curves for $w_{eff}$ model (left panel) and $w$CDM model (right panel) are shown. The values of the model parameters are fixed at $\alpha=0.444$ and $n=2.907$ for $w_{eff}$ model and $w=−0.981$ and $\Omega_{m0}=0.296$ for $w$CDM model.}}
\label{fig:delc}
\end{figure}

Critical density of contrast ($\delta_c$) is another important quantity in the study of spherical collapse. The critical density contrast $\delta_c$ is the value of linear density constant at the time of collapse. The redshift of collapse changes with the change of initial values of $\delta$ and $\delta'$. Using the same initial conditions in linear density contrast equation (eq(\ref{del_lin_a})) the value of critical density contrast can be obtained. Thus changing the initial conditions, the critical density contrast $(\delta_c)$ can be achieved as a function of redshift ($z_c$). In the present analysis, only the initial value of $\delta$ is varied, the initial value of $\delta'$ is kept unchanged. The curves of $\delta_c(z_c)$ for the $w_{eff}$ and $w$CDM model are shown in figure \ref{fig:delc}. The plots in figure \ref{fig:delc}  clearly show that the $\delta_c(z_c)$ follows the same evolution for both the reconstructed dark energy models.  The $\delta_c(z_c)$ is essential in the study of spherical collapse and number count of dark matter halos along redshift. 

\section{Halo mass function and cluster number count}
\label{clustercount}

In this section, the number count of collapsed objects or dark matter halos of a given mass range is studied. The baryonic  matter and dark matter follow the same distribution. Thus the observed distribution of galaxy clusters indicates the distribution of dark matter halos in the universe. In the present study, two different mass functions are used to obtain the cluster number count along the redshift, namely the Press-Schechter mass function \cite{Press:1973iz} and the Sheth-Tormen mass function \cite{Sheth:1999mn}. The Sheth-Torman mass function is a generalization of the Press-Schechter mass-function. The mass functions are formulated based on the assumption of Gaussian distribution of matter density field.
The comoving number density of collapsed objects (dark matter halos) along redshift $z$ having mass range $M$ to $M+dM$ is given as,
\be
\frac{dn(M,z)}{dM}=-\frac{\rho_{m0}}{M}\frac{d\ln{\sigma(M,z)}}{dM}f(\sigma(M,z)),
\ee
where $f(\sigma)$ is the mass function. The Press-Schechter \cite{Press:1973iz} mass function is given as,
\be
f_{PS}(\sigma)=\sqrt{\frac{2}{\pi}}\frac{\delta_c(z)}{\sigma(M,z)}\exp{\left[-\frac{\delta^2_c(z)}{2\sigma^2(M,z)} \right]}.
\label{fPS}
\ee
The $\sigma(M,z)$ is the corresponding rms density fluctuation in a sphere of radius $r$ enclosing a mass M. Linearized growth factor is defined as $g(z)=\delta(z)/\delta(0)$ where $\delta(z)$ is linear density contrast. The rms of density fluctuation within a sphere of radius $r_8=8h^{-1}$Mpc is written as $\sigma_8$ and it is measured directly from cosmological observations.  
Finally the rms density fluctuation $\sigma(M,z)$ is expressed as,
\be
\sigma(z,M)=\sigma_8\left(\frac{M}{M_8}\right)^{-\gamma/3}g(z),
\ee
where $M_8=6\times10^{14}\Omega_{m0}h^{-1}M_{\odot}$ is the mass within a sphere of radius $r_8$ and the $M_{\odot}$ is the solar mass. The $\gamma$ is expressed as
\be
\gamma=(0.3\Omega_{m0}h+0.2)\left[2.92+\frac{1}{3}\log{\left(\frac{M}{M_8}\right)}\right].
\ee
The effective number of collapsed objects in mass range $M_i<M<M_s$ per redshift and square degree yield as,
\be
{\mathcal N}(z)=\int_{1deg^2}d\Omega\left( \frac{c}{H(z)}\left[\int_0^z\frac{c}{H(x)}dx\right]^2\right)\int_{M_i}^{M_s}\frac{dn}{dM}dM.
\label{numbercount_eq}
\ee
\begin{figure}[tb]
\begin{center}
\includegraphics[angle=0, width=0.4\textwidth]{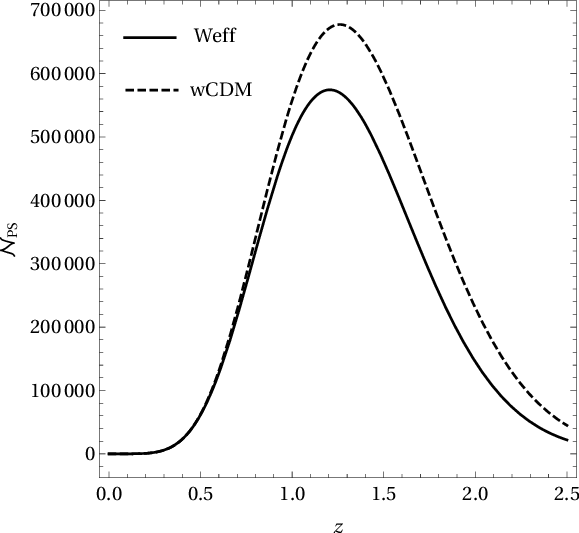}
\includegraphics[angle=0, width=0.4\textwidth]{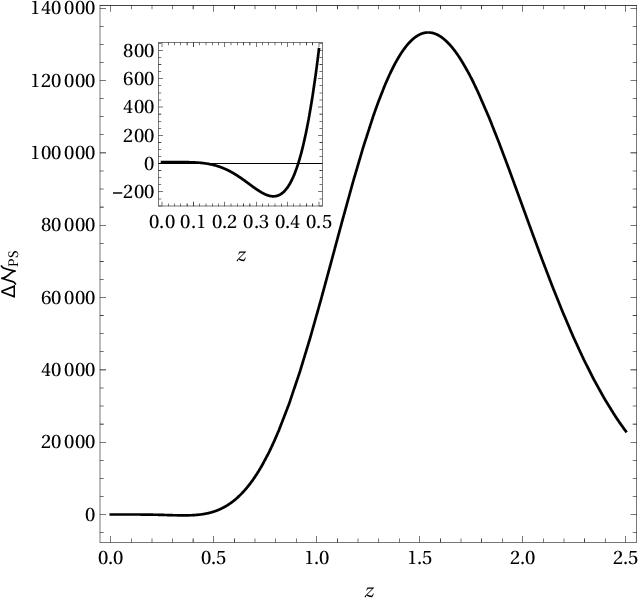}
\end{center}
\caption{{\small The left panel shows the cluster number count plots as functions of redshift obtained using the Press-Schechter mass function formula. The plots for the reconstructed $w_{eff}$ model (solid curve) and $w$CDM model (dashed curve) are shown. The right panel  shows the difference of cluster number count $\Delta{\mathcal N}_{PS}=({\mathcal N}_{wCDM}-{\mathcal N}_{weff})$. The inset of the right panel shows the $\Delta{\mathcal N}$ in very low redshift range. The values of the parameters are fixed at $\alpha=0.444$ and $n=2.907$ for $w_{eff}$ model and $w=−0.981$ and $\Omega_{m0}=0.296$ for $w$CDM.}}
\label{fig:Nplot_PS}
\end{figure}
\begin{figure}[tb]
\begin{center}
\includegraphics[angle=0, width=0.4\textwidth]{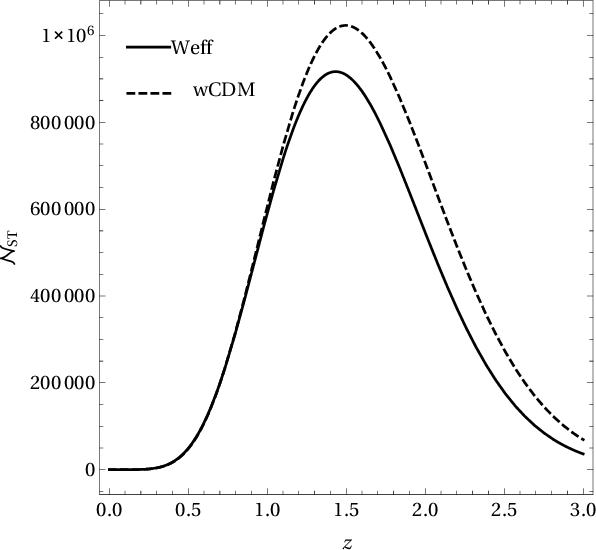}
\includegraphics[angle=0, width=0.4\textwidth]{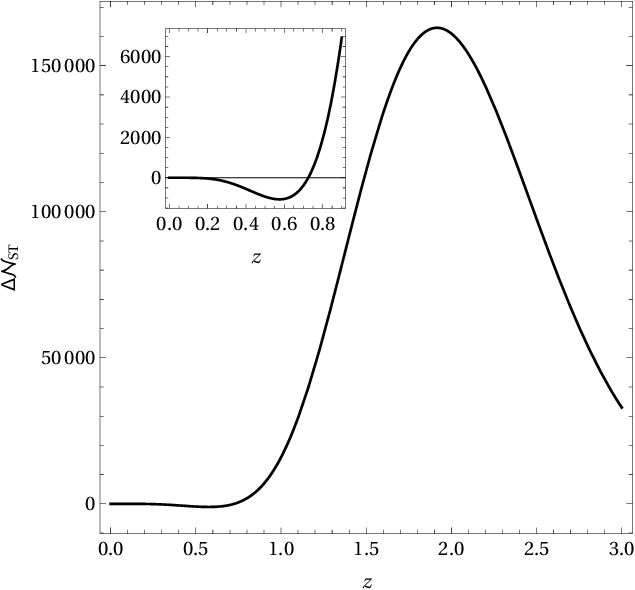}
\end{center}
\caption{{\small The left panel shows the cluster number count plots as functions of redshift obtained using the Sheth-Tormen mass function formula. The plots for the reconstructed $w_{eff}$ model (solid curve) and $w$CDM model (dashed curve) are shown. The right panel  shows the difference of cluster number count $\Delta{\mathcal N}_{ST}=({\mathcal N}_{wCDM}-{\mathcal N}_{weff})$. The inset of the right panel shows the $\Delta{\mathcal N}$ in very low redshift range. The values of the parameters are fixed at $\alpha=0.444$ and $n=2.907$ for $w_{eff}$ model and $w=−0.981$ and $\Omega_{m0}=0.296$ for $w$CDM.}}
\label{fig:Nplot_ST}
\end{figure}

The number count of collapsed object along the redshift is studied for the $w_{eff}$ model and $w$CDM model using equation (\ref{numbercount_eq}). The values of Hubble constant $H_0$ and the $\sigma_8$ are fixed at the Planck-$\Lambda$CDM measurements as $H_0=67.66\pm 0.42$ km s$^{-1}$Mpc$^{-1}$, $\sigma_8=0.8102\pm 0.0060$ (CMB power spectra+CMB lensing+BAO) \cite{Planck:2018vyg}. As the models are consistent with $\Lambda$CDM at 1$\sigma$ level, the $\Lambda$CDM estimated values of these parameters can be safely used in the present context.

The number distribution of collapsed objects, obtained for the Press-Schechter mass function formula (equation \ref{fPS}) are shown in figure \ref{fig:Nplot_PS} (left panel of figure \ref{fig:Nplot_PS}). The difference of number distribution for these two models  $\Delta{\mathcal N}_{PS}(z)=({\mathcal N}_{wCDM}(z)-{\mathcal N}_{w_{eff}}(z))$, are also shown in the right panel figure \ref{fig:Nplot_PS}. The result shows that cluster  number for the $w_{eff}$ model is lower than the cluster number for $w$CDM model at redshift $z>0.45$. At redshift $z<0.45$, the cluster number count for $w_{eff}$ model is found to be slightly higher than that of $w$CDM model (inset of the right panel of figure \ref{fig:Nplot_PS}).

A general nature of cluster number distribution along redshift is obtained for the Press-Shechter mass function. But discrepancy arises as it predicts higher number of galaxy clusters or dark matter halos at low redshift and lower number of halos at higher redshift compared to the results obtained in the simulation of large scale structure formation \cite{Reed:2006rw}. A modified mass function formula is proposed by Sheth and Torman \cite{Sheth:1999mn} which can successfully alleviate this discrepancy. The Sheth-Torman mass function is given as

\be
f_{ST}(\sigma)=A\sqrt{\frac{2}{\pi}}\left[1+\left( \frac{\sigma^2(M,z)}{a\delta^2_c(z)}\right)^p\right]\frac{\delta_c(z)}{\sigma(M,z)}\exp{\left[-\frac{a\delta_c^2(z)}{2\sigma^2(M,z)}\right]}.
\label{fST}
\ee
Three new parameters $(a,p,A)$  are introduced in the Sheth-Tormen mass function formula (eq.\ref{fST}). It can be easily checked that if the values of the new parameters are fixed as $(1,0,\frac{1}{2})$, the Sheth-Tormen mass function formula becomes the Press-Schechter mass function formula. In the present analysis, the values of the new parameters in the Sheth-Tormen  mass function is fixed according to the results obtained in the simulation of halo formation discussed in \cite{Reed:2006rw}. The parameters $(a,p,A)$ values are fixed at $(0.707,0.3,0.322)$. The cluster number count along redshift obtained for the Sheth-Tormen mass function formula (equation \ref{fST}) for these two reconstructed dark energy models are shown in figure \ref{fig:Nplot_ST}. The difference of cluster number count for these two models in case of Sheth-Tormen mass function formula $\Delta{\mathcal N}_{ST}=({\mathcal N}_{wCDM}-{\mathcal N}_{weff})$ are also shown in the right panel of figure \ref{fig:Nplot_ST}. Similar to the results obtained for the Press-Schechter mass function, the cluster number for Sheth-Tormen mass function is also found to be lower for $w_{eff}$ model compared to $w$CDM model in redshift range $z>0.7$ and in the redshift range $z<0.7$ the cluster number is slightly higher for $w_{eff}$ model compared to the $w$CDM model (right panel of figure \ref{fig:Nplot_ST}).

\begin{figure}[tb]
\begin{center}
\includegraphics[angle=0, width=0.4\textwidth]{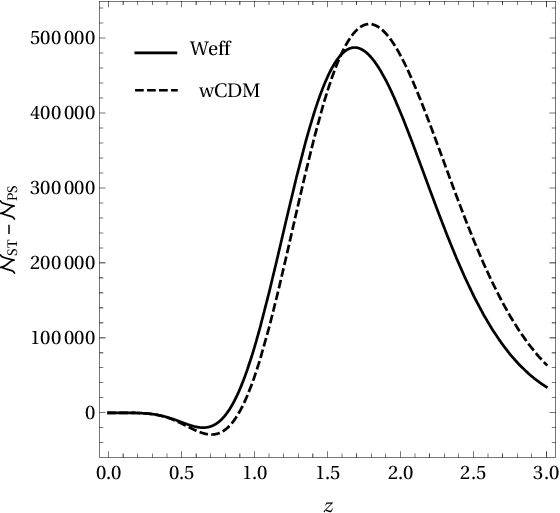}
\includegraphics[angle=0, width=0.4\textwidth]{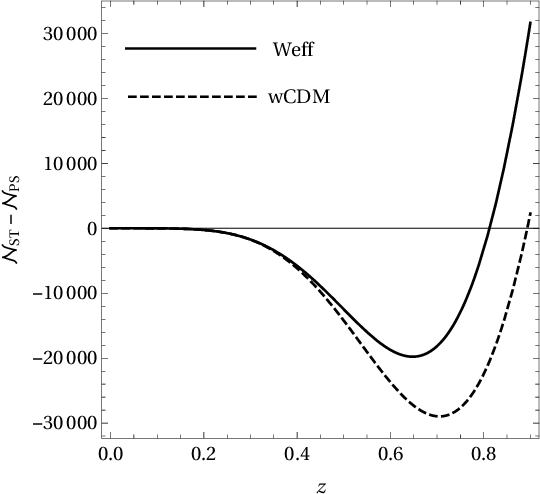}
\end{center}
\caption{{\small Plots show the difference between cluster number along redshift obtained for Press-Schechter and Sheth-Torman mass function formulae $({\mathcal N}_{ST}-{\mathcal N}_{PS})$ for reconstructed $w_{eff}$ model (solid curve) and $w$CDM model (dashed curve). The left panel shows the plot in the redshift range $0<z<3$ and the right panel shows the same plot in the redshift range $0<z<0.9$. The values of the parameters are fixed at $\alpha=0.444$ and $n=2.907$ for $w_{eff}$ model and $w=−0.981$ and $\Omega_{m0}=0.296$ for $w$CDM.}}
\label{fig:NST_NPS}
\end{figure}
  
The difference of cluster number count obtained for Sheth-Torman mass function and Press-Schechter mass function (${\mathcal N}_{ST}-{\mathcal N}_{PS}$) are also studied in the preset context. The plot of (${\mathcal N}_{ST}-{\mathcal N}_{PS}$) for these two dark energy models are shown in figure \ref{fig:NST_NPS}. It is already mentioned that the Press-Schechter mass function predicts higher cluster number at very low redshift and lower cluster number at higher redshift compared to the results obtained in simulations of large scale structure and that is why the Sheth-Torman mass function is introduced in the study of large scale structure formation. The same fact is reflected in the results obtained in  the present study. The Sheth-Tormen mass function produces higher cluster number at high redshift range ($z>0.9$) compared to the Press-Schechter mass function for both the models. At low redshift range ($z<0.8$), the cluster number is slightly higher for Press-Schechter mass function compared to the cluster number for Sheth-Tormen mass function for both the models.

\section{Conclusion}
\label{conclu}

In the present work, the nonliner evolution of matter overdensity and the formation of cosmic large scale structure is studied for two reconstructed dark energy models, namely the reconstructed $w_{eff}$ model and the $w$CDM model. Semi-analytic prescription of spherical collapse is adopted to study the evolution of matter density perturbation. The dark energy component is considered to be exclusively non-clustering. These two models are found to be degenerate at back ground and linear perturbation level. Even the evolution of non-linear perturbation (figure \ref{fig:dela_plots}) does not have any significant difference for these two models. It is also observed for both the models that the evolution of nonlinear matter perturbation is more effected by the variation of parameter values compared to linear evolution (figure \ref{fig:NL_dela_weff}, \ref{fig:L_dela_weff}).

Further, the critical density at collapse ($\delta_c$) has been studied. The $\delta_c$  is strongly dependent on the numerical infinity of nonlinear matter density contrast at the collapse. In the present work, the nonlinear matter density contrast ($\delta$) at the starting of collapse is considered to be $\delta>10^{8}$ which is important for proper convergence of the results \cite{Pace:2017qxv}. The value of redshift of collapse is changed by changing the initial conditions, i.e. the initial values of $\delta$ and $\delta'$,  whiling solving nonlinear equation of density contrast (eq. \ref{del_nonlin_a}). It is imperative to mention in this context that  fixing proper initial value of the scale factor $a_{i}$ is important  to obtain the asymptotic flatness of $\delta_c$.
Pace {\it et al}\cite{Pace:2013pea} have shown that $a_i$ should be $a_i\leq 10^{-5}$ to obtain the appropriate nature of $\delta_c$. So in the present analysis, the initial conditions are fixed at $a_i=10^{-5}$. The plots of $\delta_c(z)$ are shown in figure \ref{fig:delc}. The plots of $\delta_c$ show very similar pattern for these two models. The $\delta_c$ shows an asymptotic flatness toward the value 1.689 for both the models.

The number of dark matter halos or the galaxy clusters along redshift has also been studied for these two dark energy models. As already mentioned that in the study of cluster number count two different halo mass function formulas, namely the Press-Schechter and Sheth-Tormen mass functions, are adopted. In case of Press-Schechter mass function, it is observed that the cluster number count is substantially lower in case of the $w_{eff}$ model compared to that of $w$CDM at higher redshift ($z>0.45$) . At very low redshift ($z<0.45$) the cluster number is slightly higher for the $w_{eff}$ model (figure \ref{fig:Nplot_PS}). Similar pattern of cluster number count has been observed in case of Sheth-Torman mass function also. In case of Sheth-Torman mass function, the cluster number is higher for $w$CDM model than the $w_{eff}$ model at redshift $z>0.7$ and in redshift range $0<z<0.7$, it is slightly higher for $w_{eff}$ model (figure \ref{fig:Nplot_ST}).  It has already been mentioned that the Press-Schechter mass function predicts higher abundance of cluster number at very low redshift and lower abundance at high redshift compared to the observed number of galaxy clusters along redshift. The Sheth-Torman mass function has been introduced to alleviate this problem. Similar result has been observed in the present study also. The cluster number is found to be higher for Sheth-Torman mass function than that of Press-Schechter mass function at redshift $z>0.8$ for $w_{eff}$ model and at redshift $z>0.9$ for $w$CDM model. On the other hand in the redshift range $0<z<0.8$, the cluster number count for Press-Schechter mass function is slightly  higher than the cluster number count for Sheth-Torman mass function (figure \ref{fig:NST_NPS}).

It is already mentioned that these two reconstructed dark energy models adopted in the present study are highly degenerate at background and linear perturbation level. Even the non-linear evolution of matter perturbation have not shown any significant difference for these two models. But when the number of dark matter halos along the redshift has been studied using the formalism of spherical collapse, the degeneracy is broken. The cluster number count has indicated a significant difference between these two models. It is unequivocal from the present work that the study of nonlinear collapse of matter overdensity and cluster number count using spherical collapse is a powerful method to eliminate the degeneracy among dark energy models.  The number of collapsed objects or the dark matter halos, obtained for the theoretical models, can be confronted with observed numbers distribution of galaxy clusters in the universe. Thus we can obtain better constraints on dark energy models using present and future observations of cosmic large scale structure. The study of nonlinear collapse and number count of collapsed objects in different dark energy models using the spherical collapse model or by using N-body simulation would be useful for validation of any dark energy model based on future observations like South Pole Telescope, eROSITA etc.

\vskip 1.0 cm

\section*{Acknowledgment}
The author acknowledges the financial support from the Science and Engineering Research Board (SERB), Department of Science and Technology, Government of India through National Post-Doctoral Fellowship (NPDF, File no. PDF/2018/001859). The author would like to thank Prof. Anjan A. Sen for useful discussions and suggestions. The author would also like to thank the anonymous referee for comments and suggestions that lead to a substantial improvement of the article.

\end{document}